\documentclass[%
superscriptaddress,
preprint,
 amsmath,amssymb,
 aps,
]{revtex4-2}

\usepackage{graphicx}
\usepackage{dcolumn}
\usepackage{bm}
\usepackage{braket}
\usepackage{cancel}
\usepackage{mathtools}
\usepackage{color}
\usepackage{lineno}

\begin{document}


\title{Quantum arbitrary waveform generator}

\author{Kan Takase}
\affiliation{%
 Department of Applied Physics, School of Engineering, The University of Tokyo, 7-3-1 Hongo, Bunkyo-ku, Tokyo 113-8656, Japan
}%
\affiliation{%
 Optical Quantum Computing Research Team, RIKEN Center for Quantum Computing, 2-1 Hirosawa, Wako, Saitama 351-0198, Japan
}%

\author{Akito Kawasaki}
\affiliation{%
 Department of Applied Physics, School of Engineering, The University of Tokyo, 7-3-1 Hongo, Bunkyo-ku, Tokyo 113-8656, Japan
}%

\author{Byung Kyu Jeong}
\affiliation{%
 Department of Applied Physics, School of Engineering, The University of Tokyo, 7-3-1 Hongo, Bunkyo-ku, Tokyo 113-8656, Japan
}%

\author{Takahiro Kashiwazaki}
\author{Takushi Kazama}
\author{Koji Enbutsu}
\author{Kei Watanabe}
\author{Takeshi Umeki}
\affiliation{%
NTT Device Technology Labs, NTT Corporation, 3-1 Morinosato Wakamiya, Atsugi, Kanagawa 243-0198, Japan
}%

\author{Shigehito Miki}
\affiliation{%
Advanced ICT Research Institute, National Institute of Information and Communications Technology, 588-2 Iwaoka, Nishi-ku, Kobe, Hyogo 651-2492, Japan
}%
\affiliation{%
Graduate School of Engineering, Kobe University, 1-1 Rokkodai-cho, Nada-ku, Kobe, Hyogo 657-0013, Japan
}%

\author{Hirotaka Terai}
\author{Masahiro Yabuno}
\author{Fumihiro China}
\affiliation{%
Advanced ICT Research Institute, National Institute of Information and Communications Technology, 588-2 Iwaoka, Nishi-ku, Kobe, Hyogo 651-2492, Japan
}%

\author{Warit Asavanant}
\author{Mamoru Endo}
\affiliation{%
 Department of Applied Physics, School of Engineering, The University of Tokyo, 7-3-1 Hongo, Bunkyo-ku, Tokyo 113-8656, Japan
}%
\affiliation{%
 Optical Quantum Computing Research Team, RIKEN Center for Quantum Computing, 2-1 Hirosawa, Wako, Saitama 351-0198, Japan
}%

\author{Jun-ichi Yoshikawa}
\affiliation{%
 Optical Quantum Computing Research Team, RIKEN Center for Quantum Computing, 2-1 Hirosawa, Wako, Saitama 351-0198, Japan
}%

\author{Akira Furusawa}
\email{akiraf@ap.t.u-tokyo.ac.jp}
\affiliation{%
 Department of Applied Physics, School of Engineering, The University of Tokyo, 7-3-1 Hongo, Bunkyo-ku, Tokyo 113-8656, Japan
}%
\affiliation{%
Optical Quantum Computing Research Team, RIKEN Center for Quantum Computing, 2-1 Hirosawa, Wako, Saitama 351-0198, Japan
}%

\date{\today}


\begin{abstract}
Controlling the waveform of light is the key for a versatile light source in classical and quantum electronics. Although pulse shaping of classical light is a mature technique and has been used in various fields, more advanced applications would be realized by a light source that generates arbitrary quantum light with arbitrary temporal waveform. We call such a device a {\it quantum arbitrary waveform generator} (Q-AWG). The Q-AWG must be able to handle versatile quantum states of light, which are fragile. Thus, the Q-AWG requires a radically different methodology from classical pulse shaping. In this paper, we invent an architecture of Q-AWGs that can operate semi-deterministically at a repetition rate over GHz in principal. We demonstrate its core technology via generating highly non-classical states with waveforms that have never been realized before. This result would lead to powerful quantum technologies based on Q-AWGs such as practical optical quantum computing.
\end{abstract}

\maketitle


\section*{\label{Introduction}Introduction}
\begin{figure}[htbp]
	\begin{center}
		\includegraphics[bb= 0 0 2550 640,clip,width=\textwidth]{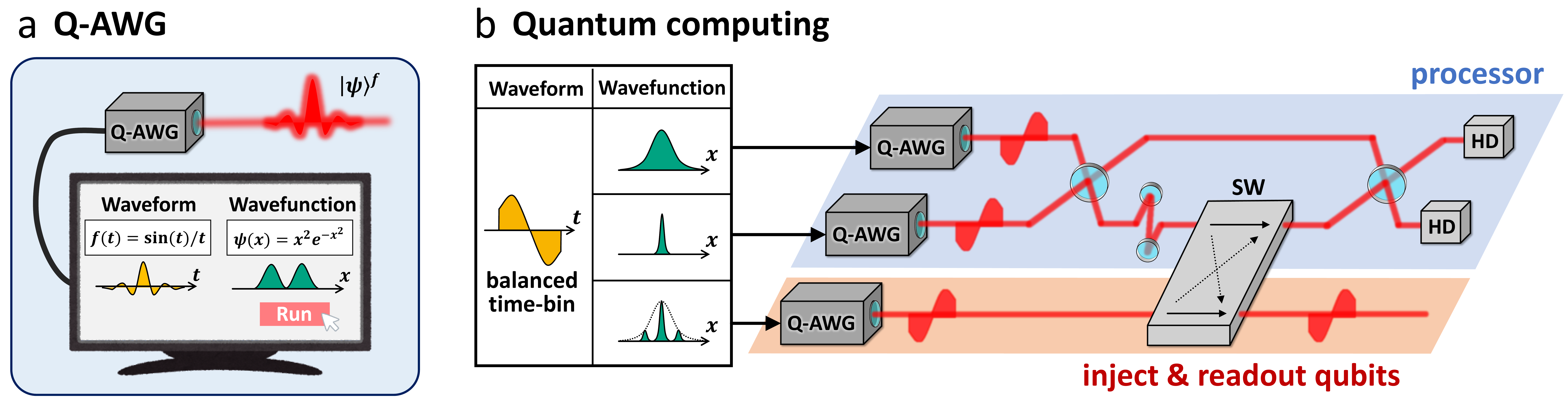}
	\end{center}
	\caption{The concept and application of a {\it quantum arbitrary waveform generator} (Q-AWG). (a) Operation image. A user specifies the  temporal waveform and wavefunction of the output quantum light. (b) Scalable quantum computing using Q-AWGs. Quantum states are efficiently multiplexed in time domain by utilizing a balanced time-bin waveform. Qubits that have non-Gaussian wavefunctions are injected into the processor by an optical switch (SW). Computing is carried out in a measurement-based way using homodyne detectors (HDs). The result of the computing is readout from the processor by the SW.}
\label{Fig:intro}
\end{figure}

The development of light sources is the key to innovation in science and technology. One useful light source is known as an arbitrary waveform generator (AWG) \cite{Cundiff2010,Weiner2011}, a device that outputs laser light with an arbitrary temporal waveform. An AWG has been used for coherent control of light-matter interaction \cite{K.-K.2022,Stowe2006,Morichika2019}, optical communication \cite{Geisler2009,Geisler2011}, spectroscopy \cite{ThorpeMichael2022,Cundiff2013,Fuller2014}, and manipulation of quantum entanglement \cite{Arzani2018,Nokkala2018}. The AWG, however, only deals with classical light, and thus is not suitable for applications that explicitly utilize quantum nature of light such as quantum computing \cite{OBrienJeremy2007,Takeda2019}, quantum networking \cite{Gisin2002,Gisin2007,Kimble2008}, and quantum sensing \cite{Abadie2011}. To get rid of this limitation, we propose a concept of a {\it quantum arbitrary waveform generator} (Q-AWG), a device that outputs an arbitrary quantum state of light with an arbitrary waveform. Figure \ref{Fig:intro}(a) shows a Q-AWG which can outputs desired quantum light by specifying a wavefunction $\psi(x)$ and a temporal waveform $f(t)$. Here, $\psi(x)$ characterizes the output quantum state and $f(t)$ characterizes the temporal distribution of the quantum state. Realization of a Q-AWG directly leads to various quantum applications. As an example, Fig. \ref{Fig:intro}(b) is scalable measurement-based quantum computing that multiplexes quantum states of light in time domain. In this scheme, it is effective to use a balanced time-bin waveform \cite{Yoshikawa2016,LarsenMikkel2019,Warit2019,Larsen2021,Asavanant2021a,Shuntaro2021,Yutaro2021}. By outputting various quantum states with this temporal waveform from Q-AWGs, large-scale, universal, and fault-tolerant optical quantum computing can be performed. The Q-AWG would also play a central role in quantum networking and quantum sensing by generating other waveforms such as exponentially rising \cite{Stobinska2009,Khalili2010,Nunn2007,Yoshikawa2013} or temporal-grid waveforms \cite{Fabre2021,Bruschi2021}.

As shown in Fig. \ref{Fig:intro}(b), quantum states can have both Gaussian and non-Gaussian wavefunctions, where those states are called Gaussian and non-Gaussian states, respectively. If all we need is Gaussian states, realizing a Q-AWG is not demanding \cite{Yoshikawa2016,Arzani2018,Nokkala2018}. In particular, control of coherent states, which are loss-tolerant Gaussian states, is well developed by conventional AWG \cite{Cundiff2010,Weiner2011}. In many quantum applications, however, non-Gaussian states are an essential factor \cite{Vittorio2022,Ohliger2010,Mari2012}. Generation of non-Gaussian states and controlling its temporal waveform require a completely different methodology from Gaussian states, making it difficult to realize a Q-AWG. Although arbitrary non-Gaussian wavefunction can be realized via heralding state-generation scheme in principle \cite{Yukawa2013}, the methodology for realizing arbitrary temporal waveform has remained elusive. Thus far, waveform engineering of non-Gaussian states has been discussed only in specific heralding methods. This has resulted in the demonstration of limited types of non-Gaussian states with several simple temporal waveforms in continuous-wave (CW) \cite{Ogawa2016,Asavanant2017,Konno2021,Takeda2012,Takase2019} and pulsed regime \cite{Costanzo2016,Ra2020}.

In this paper, we propose an architecture of Q-AWGs and demonstrate its core technology --- generation of an arbitrary non-Gaussian state with an arbitrary waveform. Our idea is based on the implementation of the heralding scheme using broadband CW quantum light sources. By doing so, the temporal waveforms of the non-Gaussian states are directly decided by the impulse response of optical filters in the system. By incorporating timing control technique \cite{Wang2017,Pittman2002,Kaneda2015,Shuntaro2021}, high-success-rate heralding methods \cite{Tzitrin2020,Takase2021}, and THz-bandwidth quantum light source \cite{Kashiwazaki2021b,Takase2022} into the basic idea, the Q-AWG can output states semi-deterministically with a repetition rate over GHz. As a demonstration, we generate non-Gaussian states with time-bin and balanced time-bin waveforms which have never been realized before. This result would lead to practical quantum computing shown in Fig. \ref{Fig:intro}(b), and to realization of the Q-AWG --- an ultimate quantum light source.

\section*{\label{Results}Results}

\begin{figure}[htbp]
	\begin{center}
		\includegraphics[bb= 0 0 1480 730,clip,width=\textwidth]{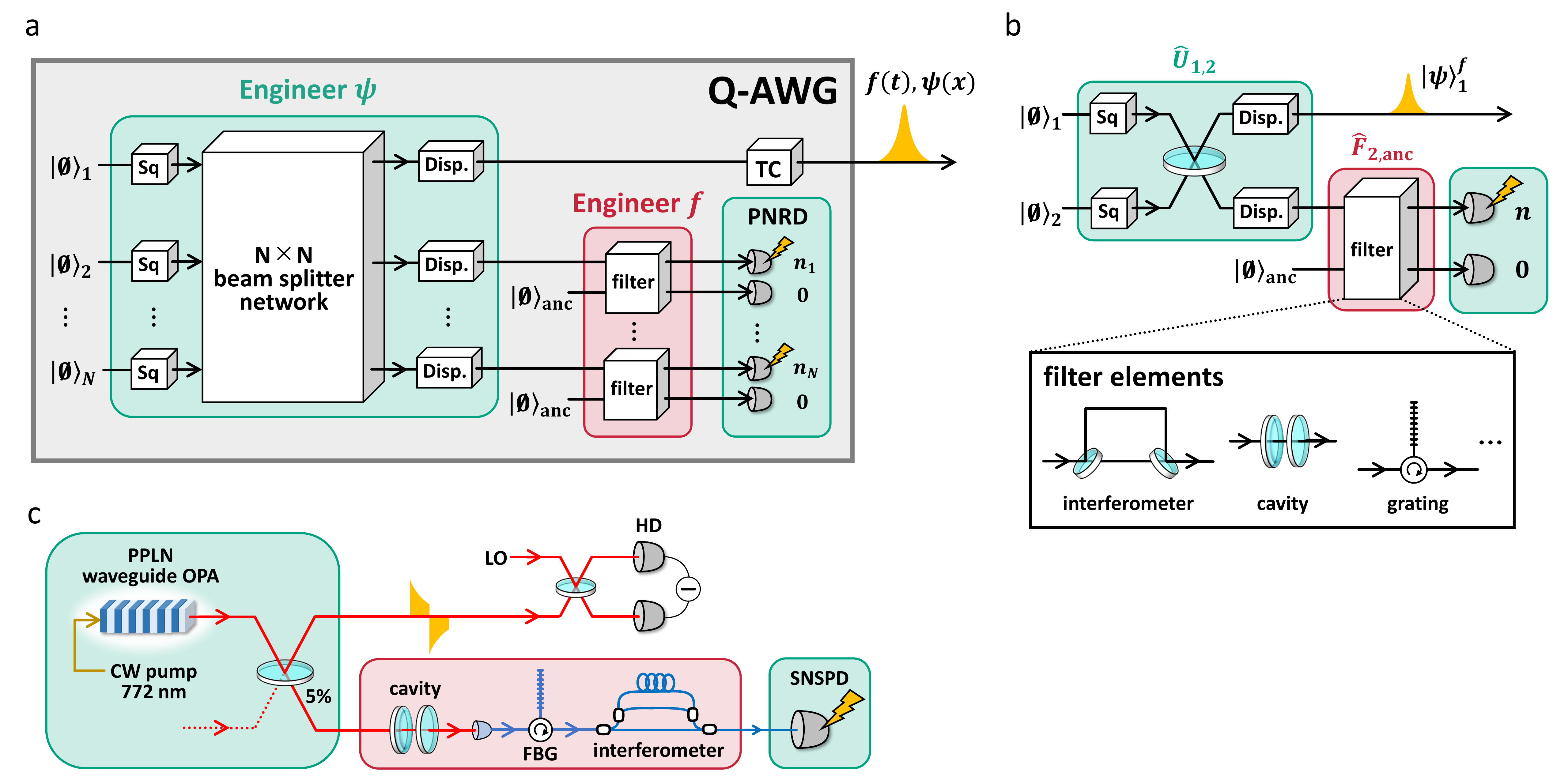}
	\end{center}
	\caption{The diagram of the proposed Q-AWG and setup of the experiment. (a) The architecture of the Q-AWG. Arbitrary non-Gaussian states with arbitrary temporal waveform are generated by heralding scheme. A timing controller (TC) adjusts the emission timing of the state. CW, Continuous Wave; PNRD, Photon Number Resolving Detector. (b) The detail of waveform engineering in 2-channel case. The experiment we conduct is the 2-channel case with the TC omitted. (c) Experimental setup. PPLN, periodically poled LiNb${\rm O_3}$; OPA, Optical
Parametric Amplifier; LO, Local Oscillator; HD, Homodyne Detector; FBG, Fiber Bragg Grating;
SNSPD, Superconducting Nanostrip Photon Detector.}
\label{Fig:theory_fig4}
\end{figure}

{\bf Architecture and operating principle.}
First, we introduce an architecture of the Q-AWG shown in Fig. \ref{Fig:theory_fig4} (a). The Q-AWG generates a non-Gaussian state via the heralding scheme, where photon number measurement is performed on a subsystem of a Gaussian state defined on multiple channels \cite{Yukawa2013,Tzitrin2020}. In principle, we can realize arbitrarily wavefunction $\psi (x)$ of the generated state by changing the initial Gaussian state and the number of photons detected \cite{Yukawa2013}. Let us suppose the initial Gaussian state is broadband CW quantum light. Then, we can realize arbitrary temporal waveform $f(t)$ of the generated state by inserting passive linear filters with a impulse response $g(t)\propto f(-t)$ before the photon detectors. Although heralding is a probabilistic scheme, a timing controller such as continuously tunable delay \cite{Wang2017} or quantum memory \cite{Pittman2002,Kaneda2015,Shuntaro2021} can make it semi-deterministic. One advantage of our method is we supposes a general Gaussian initial state. Thanks to this, our method can be applied to recently developed efficient heralding methods \cite{Tzitrin2020,Takase2021} with a success probability around $10^{-3}$ to $10^{-2}$. By incorporating these efficient heralding methods and Thz-bandwidth quantum light source \cite{Kashiwazaki2021b,Takase2022} into our method, where the bandwidth is roughly corresponds to the trial rate of state generation, the Q-AWG in Fig. \ref{Fig:theory_fig4} (a) can operate semi-deterministically at a repetition rate over GHz.

Next, we introduce the working principle of the Q-AWG in a two-channel case shown in Fig. \ref{Fig:theory_fig4} (b). We suppose continuous-wave (CW) squeezed light in channel 1 and 2 interferes at a beam splitter and each channel is displaced. We express this initial Gaussian state as $\hat{U}_{1,2}\ket{\cancel{0}}_1\ket{\cancel{0}}_2\ket{\cancel{0}}_{{\rm anc}}$, where $\ket{\cancel{0}}_j$ is a multimode vacuum state of channel $j$ satisfying $\hat{a}_j(t)\ket{\cancel{0}}_j=0$ for all $t$. An passive filter consisting of linear optics, which has an ancillary channel in quantum description \cite{Shimazu2021}, is inserted in channel 2. We perform photon number measurement on the filter outputs. When we detect $n$ photons on channel 2 at $t=0$ and detect no photon on the ancillary channel, the heralded state on channel 1 is given by 
\begin{eqnarray}\label{eq:continuous_state}
\ket{\Psi}_1\propto{}_{{\rm anc}}\bra{\cancel{0}}{}_2\bra{\cancel{0}}\hat{a}_2(t=0)^n\hat{F}_{2,{\rm anc}}\hat{U}_{1,2}\ket{\cancel{0}}_1\ket{\cancel{0}}_2\ket{\cancel{0}}_{{\rm anc}},
\end{eqnarray}
where $\hat{F}_{2,{\rm anc}}$ is a filter operator and $\hat{a}_j(t)$ is the instantaneous annihilation operator satisfying a commutation relation $\left[\hat{a}_j(t),\hat{a}_j^{\dag}(t')\right]=\delta(t-t')$.

Suppose we aim to realize a waveform $f(t)$. Generally, $f(t)$ is a complex function, but for the time being we suppose $f(t)$ is a real function. We introduce the following assumptions: (I) squeezing and beam-splitter operation are broadband and frequency-independent within the spectrum of $f(t)$, (II) the filter transmittance about channel 2 is characterized by an impulse response $g(t)\propto f(-t)\in\mathbb{R}$. The assumption on the beam splitter is usually satisfied in experiments, and thus the assumption on the squeezed light and the filter is the key. Under these assumptions, $\ket{\Psi}_1$ in Eq. (\ref{eq:continuous_state}) can be decomposed to a direct product of a single-mode state about $f(t)$ and a multimode state with orthogonal waveforms to $f(t)$ as follows,
\begin{eqnarray}
\ket{\Psi}_1&=&\ket{\psi}_1^{f}\otimes\ket{\phi}_1^{\{f_{\perp}\}}, \label{eq:decomp} \\
\ket{\psi}_1^f&\propto&\ {}_2^f\Braket{n|G}_{1,2}^f. \label{eq:nonGauss}
\end{eqnarray}
Here, $\ket{G}_{1,2}^f$ is an arbitrary Gaussian state with a waveform $f(t)$. See Methods for the derivation. $\ket{\psi}_1^{f}$ is a pure non-Gaussian state due to the effect of photon number measurement except for the cases where $\ket{G}_{1,2}^f$ is a separable state or $n=0$. Note that Eq. (\ref{eq:nonGauss}) is equivalent to the generalized heralding scheme supposing an arbitrary 2-channel Gaussian input \cite{Tzitrin2020,Takase2021}. Since we can implement a filter with an arbitrary impulse response $g(t)\in\mathbb{C}$ \cite{Ra2020,JingujiJan.,JingujiAug.}, our method can tailor arbitrary waveforms given by a real function. In addition, it is always possible to maximize the generation rate of non-Gaussian states by constructing a 100$\%$-efficiency filter that transmits the desired spectrum only to one output channel \cite{JingujiJan.,JingujiAug.}. Existence of such an efficient filter is important because low generation rate is often problematic in a heralding scheme. More general quantum states can be generated using N channels. Supposing we detect $n_j$ photons in channel $j\ (2\le j \le N)$ after the filters with $g(t)\propto f(-t)\in\mathbb{R}$, we can generate a state given by
\begin{eqnarray}
\ket{\psi}_1^f&\propto&\ {}_N^f\Bra{n_N}\cdots{}_3^f\Bra{n_3}{}_2^f\Braket{n_2|G}_{1,2,\cdots ,N}^f,
\end{eqnarray}
where $\ket{G}_{1,2,\cdots ,N}^f$ is an arbitrary Gaussian state of mode $f$ on channel 1 to N. One example of N-channel heralding utilizes Einstein-Podolsky-Rosen state, where one mode of the state is divided into N-1 channels and displaced before photon number measurement. By supposing $n_j=1\ (2\le j \le N)$, $\ket{\psi}_1^f$ can be an arbitrary superposition up to $N-1$ photons \cite{Yukawa2013}. Thus, arbitrary single-mode state can be generated when $N\rightarrow \infty$. Even so, discussing the case of general $\Ket{G}_{1,2,\cdots ,N}^f$ is important because many useful states can be generated much more efficiently in this situation \cite{Tzitrin2020,Takase2021}.

Finally, we show two possible ways to realize a waveform given by a complex function. One is to add a phase $e^{i\theta(t)}$ to the generated state; the waveform is converted from $f(t)$ to $f(t)e^{i\theta(t)}$. Another way is to use an Einstein-Podolsky-Rosen state as an input; in this case, we can realize an arbitrary complex waveform $f(t)$ by utilizing a filter with an impulse response $g(t)\propto f(-t)\in\mathbb{C}$. See Methods for the detail.

{\bf Experiment.} As a demonstration, we generate Schr\"{o}dinger cat states in time-bin and balanced time-bin waveforms. The cat states are non-Gaussian states defined as $\ket{\alpha}\pm\ket{-\alpha}$, where $\ket{\alpha}$ is a coherent state with amplitude $\alpha$. The target waveforms are given by decay rate $\Gamma$ and bin duration $\Delta t$ as follows,
\begin{eqnarray}
f_{{\rm TB}}(t) &=& \sqrt{\frac{2\Gamma}{\exp{[2\Gamma \Delta t]}-1}}\times
 \left\{
\begin{array}{ll}
0 & (t < 0, \Delta t\le t)\\
\exp{[\Gamma t]} & (t \ge 0),
\end{array}
\right. \label{eq:f_TB} \\
f_{{\rm BTB}}(t) &=& \sqrt{\frac{\Gamma}{\exp{[2\Gamma \Delta t]}-1}}\times
 \left\{
\begin{array}{ll}
0 & (t < 0,2\Delta t \le t)\\
\exp{[\Gamma t]} & (0\le t < \Delta t) \\
-\exp{[\Gamma (t-\Delta t)]} & (\Delta t\le t < 2\Delta t),
\end{array}
\right. \label{eq:f_BTB}
\end{eqnarray}
where TB and BTB denote time-bin and balanced time-bin, respectively. These functions have non-zero values only at finite time domain, and the balanced time-bin waveform has no career component because $\int f_{{\rm BTB}}(t)\ dt=0$. Balanced time-bin waveform is suitable for time-domain multiplexing of quantum states and thus intensively adopted as a computational basis in scalable optical quantum computers \cite{Yoshikawa2016,LarsenMikkel2019,Warit2019,Larsen2021,Asavanant2021a,Shuntaro2021,Yutaro2021}. Thus far, those waveforms have not been demonstrated in non-Gaussian state generation. 

We conduct photon subtraction \cite{Dakna1997} to generate the cat states by the setup shown in Fig. \ref{Fig:theory_fig4} (c). We demonstrate the case of $n=1$ and thus expect that the cat states $\ket{\alpha}-\ket{-\alpha}$ with $|\alpha|\sim 1$ are generated. In our method, a broadband CW squeezer and a filter with a desired impulse response are the key. As a broadband source, a single-pass waveguide optical parametric amplifier (OPA) is a promising candidate. In this experiment, we use a low-loss periodically poled LiNb${\rm O_3}$ waveguide OPA module \cite{Kashiwazaki2021b} we have recently developed. Our module can generate THz-bandwidth CW squeezed light at the wavelength of 1545 nm with high purity sufficient for non-Gaussian state generation \cite{Takase2022}. The squeezed light is emitted into free space, then 5$\%$ of it is tapped for the photon detection. We construct two filters with impulse response $g_{{\rm TB}}(t)\propto f_{{\rm TB}}(-t)$ and $g_{{\rm BTB}}(t)\propto f_{{\rm BTB}}(-t)$ before the photon detector. Based on the optical quantum information processors reported so far \cite{Warit2019,Asavanant2021a,Shuntaro2021,Yutaro2021}, we set the parameter in Eq. (\ref{eq:f_TB}) and (\ref{eq:f_BTB}) as $\Delta t=20$ ns and $\Gamma = 2\pi \times 8.2$ MHz. The filters consist of Mach-Zehnder interferometers, optical cavities, and fiber Gragg gratings. These elements are suitable to realize filters with various bandwidth from MHz order to THz order, and principally we can approximate arbitrary impulse response with arbitrary accuracy using those elements \cite{JingujiJan.,JingujiAug.}. We detect photons with a superconducting nanostrip photon detector (SNSPD) \cite{Miki2017}, and the timing jitter of photon detection is about 100 ps, sufficiently small for this experiment. Each click at the SNSPD heralds a cat state in the other channel. We omit the photon number measurement on the ancillary outputs of the filter because simultaneous photon detection at the outputs is rare.

{\bf Evaluation.} 
We evaluate the filters and the generated states by using homodyne detectors with 200-MHz bandwidth. Supposing $\theta_{{\rm LO}}$ is the phase of the CW local oscillator (LO) beam, we measure the instantaneous quadrature $[\hat{a}(t)e^{-i\theta_{{\rm LO}}}+\hat{a}^{\dag}(t)e^{i\theta_{{\rm LO}}}]/\sqrt{2}=\hat{x}(t)\cos{\theta_{{\rm LO}}}+\hat{p}(t)\sin{\theta_{{\rm LO}}} \coloneqq \hat{x}_{\theta_{{\rm LO}}}(t)$. First, we evaluate the impulse response of the filters. We input weak pulsed-laser light to the filters and measure the output for 1,000 times with random $\theta_{{\rm LO}}$. The duration of the input pulse is less than 1 ps. We estimate the impulse response by applying  principal component analysis \cite{Abdi2021,MacRae2012,Morin2013} to the auto-correlation of homodyne signal. The waveform of the generated states is also estimated by principal component analysis, where we measure 20,000 events for each phase $\theta_{{\rm LO}}=0,30,60,90,120,$ and 150 degree. We measure vacuum states in the same condition as a reference for the shotnoise level. The efficiency of the homodyne detector for generated state evaluation is 0.93. Experimentally generated states are always mixed states, and thus Wigner function is often used for evaluation rather than wavefunction. Especially, the mixed state is usually regarded as non-Gaussian state only when its Wigner function has negative values. Using the estimated waveform $f(t)$, we calculate the quadratures of the mode $\int f(t)\hat{x}_{\theta_{{\rm LO}}}(t)$ and reconstruct the Wigner function of the generated states by quantum state tomography \cite{Lvovsky2009}.

\begin{figure}[htbp]
	\begin{center}
		\includegraphics[bb= 0 0 2020 1110,clip,width=\textwidth]{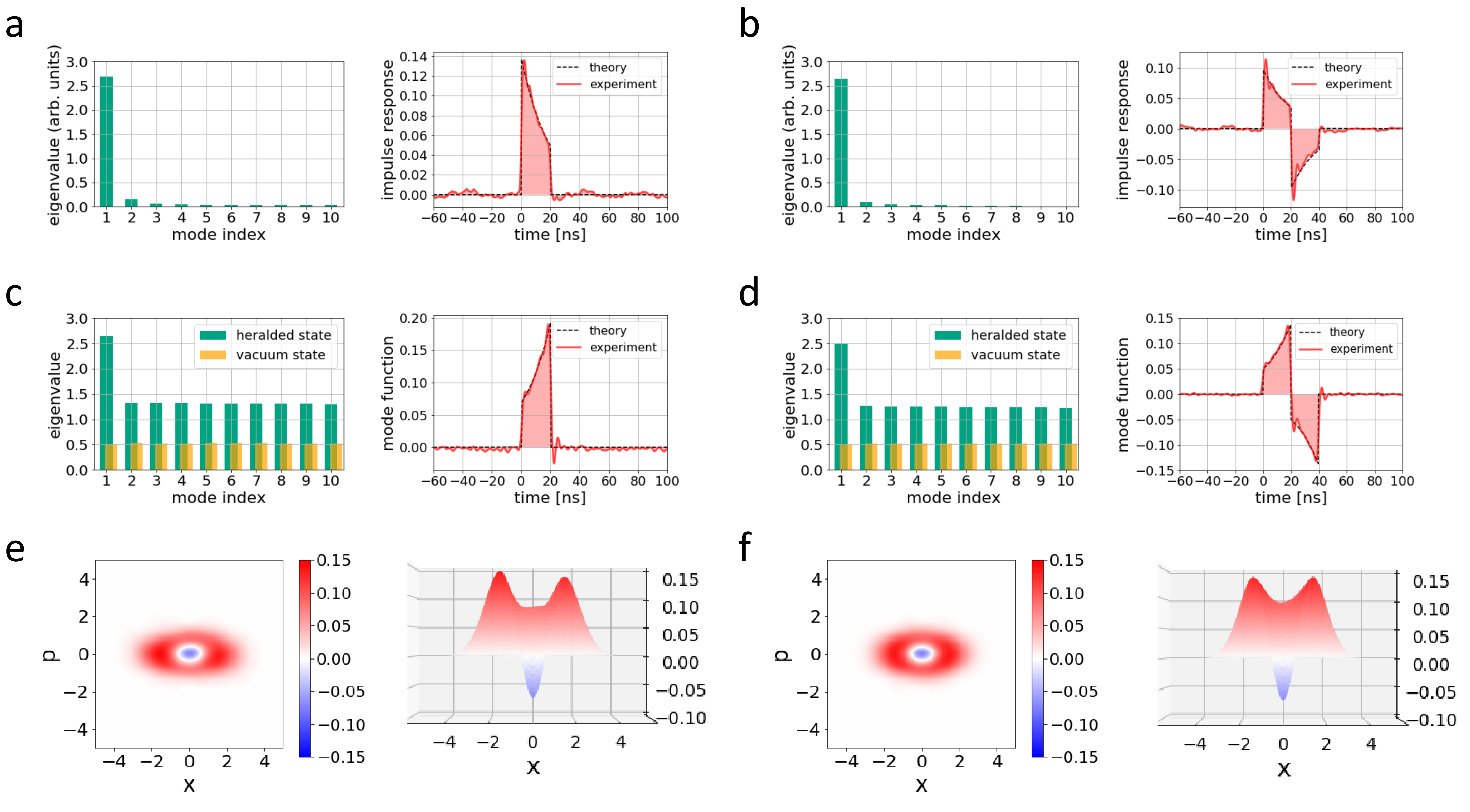}
	\end{center}
	\caption{Results of the experiment. (a) Results of principal component analysis about the filter for the time-bin waveform. The bar chart is the eigenvalues and the line chart is the first eigenfunction. (b) Similar figures to (a) about the balanced time-bin waveform. (c) Results of principal component analysis about the heralding generation of states in the time-bin waveform. The eigenvalues are the case of $\theta_{{\rm LO}}=0$ degree, and the waveform is estimated from all the measured data.  (d) Similar figures to (c) about balanced time-bin waveform. (e) Wigner function of the state generated in the estimated waveform shown in (c). (f) Wigner function of the state generated in the estimated waveform shown in (d).}
\label{Fig:result}
\end{figure}

Figure \ref{Fig:result} shows the experimental results. Figure \ref{Fig:result} (a) and (b) are the results of impulse response evaluation about time-bin and balanced time-bin waveforms, respectively. These figures show the eigenvalues and the first eigenfunction of principal component analysis. In both figures, the first eigenvalue is much larger than the other eigenvalues, indicating that the impulse response of this filter can be expressed as a real function and that the cavity and interferometer of the filter are properly controlled. The first eigenfunction is the estimated impulse response of the filter. We can observe finite time duration of the impulse response as expected. The mode match with the theoretical prediction is 0.971 for time-bin waveform and 0.954 for balanced time-bin, and thus the filters are implemented with a good accuracy. Figure \ref{Fig:result} (c) and (d) are the results of waveform evaluation of the generated quantum states. The green bars in these figures are the eigenvalues of principal component analysis of the generated states with $\theta_{{\rm LO}}=0$ degree. Corresponding eigenfunctions are the waveforms of those components. We can see that only the first eigenvalue is outstanding, indicating that single-mode non-Gaussian states are generated with the waveform given by the first eigenfunction. The orange bars are the variance of the quadrature of vacuum states with those waveforms. We estimate the waveforms of the generated states as the average of the first eigenfunctions of all $\theta_{{\rm LO}}$. The line charts in Fig. \ref{Fig:result} (c) and (d) are the estimated waveforms. Since the squeezed light is sufficiently broadband, these functions are given by the time-reversed impulse responses shown in Fig. \ref{Fig:result} (a) and (b). The mode match with the theoretical prediction is 0.958 for time-bin waveform and 0.954 for balanced time-bin. Fig. \ref{Fig:result} (e) and (f) are the Wigner functions of the states in the estimated time-bin and balanced time-bin waveforms. These states are non-Gaussian states with high non-classicality because each Wigner function has negative value $-0.070\pm 0.002$ and $-0.068\pm 0.002$ at the origin of the phase space, respectively. These states best approximate a cat state with $|\alpha|=0.94$, where the fidelities are $F = 0.564\pm 0.001$ for the time-bin waveform and $F = 0.604\pm 0.03$ for the balanced time-bin. The fidelity can get much better by improving the main imperfection of the system: loss at the waveguide OPA, inefficiency of the homodyne detector and SNSPD, and fake count of the SNSPD. From all the above, we have successfully demonstrated the core technology of the proposed Q-AWG via generating non-Gaussian states in time-bin and balanced time-bin waveforms.

\section*{Discussion}
Our achievement is twofold: first, we have proposed the concept of the Q-AWG and its architecture based on heralding scheme using CW; second, we have demonstrated the core technology of the Q-AWG, arbitrary control of the temporal waveform of non-Gaussian states, via realizing waveforms essential for practical optical quantum computing \cite{Yoshikawa2016,LarsenMikkel2019,Warit2019,Larsen2021,Asavanant2021a,Shuntaro2021,Yutaro2021}.

As a future prospect, generating of quantum states with wider bandwidth is a straightforward task. Widening the bandwidth has the following two implications. First, the wider the bandwidth, the shorter the delay lines in a filter and a timing controller, and the more compact the whole system will be. Therefore, it becomes easy to realize a complex temporal waveform by a filter consists of many elements, to make the Q-AWG semi-deterministic, and to integrate the system on a chip by the recent waveguide technology \cite{Perez2017}. Another implication is improvement of the state-generation rate. This is because the bandwidth of the generated states, that is, the bandwidth of the filter is roughly regarded as the trial rate of state generation in a heralding scheme. As fast detection of quadratures and photons are having been developed \cite{Takanashi2020,Kashiwazaki2021b,Korzh2020}, generation of non-Gaussian states with GHz-bandwidth will be possible before long. If related technologies develop further and the THz-bandwidth of the waveguide OPA is fully exploited, a semi-deterministic Q-AWG with a GHz-order repetition rate can be realized.

Multimode Q-AWGs are also an interesting research direction. By using filters with different impulse responses in a N-channel heralding scheme, our method can be generalized to arbitrary control of multimode non-Gaussian states. Multimode Q-AWGs would be used to increase the capacity of quantum information processing by multiplexing quantum states in frequency domain \cite{Roslund2014} or to generate logical states for fault-tolerant quantum computing \cite{Yoshikawa2018}.

Although our experimental result is directly relevant to optical quantum computing, the Q-AWG can also be applied to other applications such as quantum networking \cite{Stobinska2009,Khalili2010,Nunn2007,Yoshikawa2013} and quantum sensing \cite{Fabre2021,Bruschi2021}. In the future, it is quite possible that useful quantum light or powerful applications that are not yet known will be proposed. We can flexibly meet any such requirements because our method allows arbitrary control of quantum light at any wavelength. The concept of the Q-AWG and our technical achievements would be a key of optical quantum technology.

\section*{Methods}

{\bf Decomposition of CW multimode states.} 
We show the derivation of Eq. (\ref{eq:decomp}) and Eq. (\ref{eq:nonGauss}). A fundamental operator of wave packet mode $f$ is a creation operator given by
\begin{eqnarray}
\hat{a}^{\dag}[f] = \int f(t)\hat{a}^{\dag}(t)\ dt = \int \tilde{f}(\omega)\hat{\tilde{a}}^{\dag}(\omega)\ d\omega,
\end{eqnarray}
where $\hat{a}^{\dag}(t)$ and $\hat{\tilde{a}}^{\dag}(\omega)$
are the instantaneous and monochromatic creation operators satisfying commutation relations $\left[\hat{a}(t),\hat{a}^{\dag}(t')\right]=\delta(t-t'), \left[\hat{\tilde{a}}(\omega),\hat{\tilde{a}}^{\dag}(\omega')\right]=\delta(\omega-\omega')$. These operators are transformed to each other by $\hat{a}(t)=\frac{1}{\sqrt{2\pi}}\int \hat{\tilde{a}}(\omega)e^{-i\omega t}\ d\omega$ and $\hat{\tilde{a}}(\omega)=\frac{1}{\sqrt{2\pi}}\int \hat{a}(t)e^{i\omega t}\ dt$. We take a rotating frame so that the career frequency becomes $\omega=0$. The mode functions $f(t),\tilde{f}(\omega)$ satisfy
\begin{eqnarray}
\int \left|f(t)\right|^2\ dt = 1&,&\ 
\int \left|\tilde{f}(\omega)\right|^2\ d\omega = 1,\label{eq:f_norm} \\
\tilde{f}(\omega) = \frac{1}{\sqrt{2\pi}}\int f(t)e^{-i\omega t} \ dt&,&\ f(t) = \frac{1}{\sqrt{2\pi}}\int \tilde{f}(\omega)e^{i\omega t}\ d\omega.
\end{eqnarray}
Note that the mode function $f(t)$ is referred as temporal waveform in the main text. Equation (\ref{eq:f_norm}) guarantees $\left[\hat{a}[f],\hat{a}^{\dag}[f]\right]=1$. The Fock states $\left\{\ket{n_f}\right\}_{n=0}^{\infty}$ defined as the eigenstates of $\hat{a}^{\dag}[f]\hat{a}[f]$ are a complete orthonormal system of the mode $f$.

A quantum filter requires at least one ancillary input \cite{Shimazu2021}. We explicitly deal with two-mode filters, but the same argument holds for the multimode case. A passive filter operator on a channel $j$, which we denote $\hat{F}_{j,{\rm anc}}$, gives frequency-dependent unitary transformation as follows,
\begin{eqnarray}
\hat{F}_{j,{\rm anc}}^{\dag}
\begin{pmatrix}
\hat{\tilde{a}}_j(\omega) \\
\hat{\tilde{a}}_{\rm anc}(\omega) \\
\end{pmatrix}
\hat{F}_{j,{\rm anc}}
= 
\begin{pmatrix}
\tilde{g}(\omega) & \tilde{h}(\omega) \\
-\tilde{h}^*(\omega) & \tilde{g}^*(\omega) \\
\end{pmatrix}
\begin{pmatrix}
\hat{\tilde{a}}_j(\omega) \\
\hat{\tilde{a}}_{\rm anc}(\omega) \\
\end{pmatrix},
\end{eqnarray}
where $|\tilde{g}(\omega)|^2+|\tilde{h}(\omega)|^2=1$ for all $\omega$. Then Eq. (\ref{eq:continuous_state}) is given by
\begin{eqnarray}
\ket{\Psi}_1&\propto&{}_{{\rm anc}}\bra{\cancel{0}}{}_2\bra{\cancel{0}}\hat{a}_2(t=0)^n\hat{F}_{2,{\rm anc}}\hat{U}_{1,2}\ket{\cancel{0}}_1\ket{\cancel{0}}_2\ket{\cancel{0}}_{{\rm anc}}\nonumber \\
&=&{}_{{\rm anc}}\bra{\cancel{0}}{}_2\bra{\cancel{0}} \left[\frac{1}{\sqrt{2\pi}}\int\left(\tilde{g}(\omega)\hat{\tilde{a}}_2(\omega) + \tilde{h}(\omega)\hat{\tilde{a}}_{\rm anc}(\omega) \right)d\omega\right]^n \hat{U}_{1,2}\ket{\cancel{0}}_1\ket{\cancel{0}}_2\ket{\cancel{0}}_{{\rm anc}}\nonumber \\
&\propto& {}_2\bra{\cancel{0}}\hat{a}_2[N\left(g^{*R}\right)]^n \hat{U}_{1,2}\ket{\cancel{0}}_1\ket{\cancel{0}}_2,
\end{eqnarray}
where we use $\hat{F}_{2,{\rm anc}}\ket{\cancel{0}}_2\ket{\cancel{0}}_{{\rm anc}}=\ket{\cancel{0}}_2\ket{\cancel{0}}_{{\rm anc}}$ and $N(\cdot)$ denotes normalization. The Gaussian operator $\hat{U}_{1,2}$ is given by
\begin{eqnarray}
\hat{U}_{1,2} = \hat{D}_1(\alpha_1,\omega_d)\hat{D}_2(\alpha_2,\omega_d)\hat{B}_{1,2}\hat{S}_1[r_1]\hat{S}_2[r_2],
\end{eqnarray}
where CW squeezing operator $\hat{S}_j[r]$, frequency-independent beam-splitter operator $\hat{B}_{j,k}(\kappa,\nu,\mu)$, and monochromatic displacement operator $\hat{D}_j(\alpha,\omega_d)$ are defined as
\begin{eqnarray}
&&\hat{S}_j[r] = \exp{\left[\frac{1}{2}\left( \hat{P}_j^{\dag}[r]-\hat{P}_j[r] \right) \right]}, \\
&&\ \ \ \hat{P}_j^{\dag}[r] = \iint r(t_1-t_2)\hat{a}_j^{\dag}(t_1)\hat{a}_j^{\dag}(t_2)\ dt_1dt_2 = \int \tilde{r}(\omega)\hat{\tilde{a}}_j^{\dag}(\omega)\hat{\tilde{a}}_j^{\dag}(-\omega)\ d\omega, \\
&&\hat{B}_{j,k}(\kappa,\nu,\mu) = \exp{\left[i\frac{\nu}{2}\hat{L} \right]}\cdot \exp{\left[\frac{\kappa}{2}\hat{M} \right]}\cdot\exp{\left[i\frac{\mu}{2}\hat{L} \right]},\\
&&\ \ \ \hat{L} = \int \left[\hat{\tilde{a}}_j^{\dag}(\omega)\hat{\tilde{a}}_j(\omega)-\hat{\tilde{a}}_k^{\dag}(\omega)\hat{\tilde{a}}_k(\omega)\right] d\omega,\ \hat{M} = \int \left[\hat{\tilde{a}}_j^{\dag}(\omega)\hat{\tilde{a}}_k(\omega)-\hat{\tilde{a}}_k^{\dag}(\omega)\hat{\tilde{a}}_j(\omega)\right] d\omega, \\
&&\hat{D}_j(\alpha,\omega_d) = \exp{\left[ \alpha\hat{\tilde{a}}_j^{\dag}(\omega_d)-\alpha^*\hat{\tilde{a}}_j(\omega_d) \right]}.
\end{eqnarray}
From the time-reversal symmetry $r(t)=r(-t)$, $\tilde{r}(\omega)$ satisfies $\tilde{r}(\omega)=\tilde{r}(-\omega)$. 

For arbitrary wave packet mode $f$, the operators $\hat{B}_{j,k}(\kappa,\nu,\mu)$ and $\hat{D}_j(\alpha,\omega_d)$ can be decomposed to direct products of operators about mode $f$ and orthogonal modes. We can always define a complete orthonormal system $\left\{ \tilde{f}_{(l)}(\omega) \right\}_{l=1}^{\infty}$ ,where $\tilde{f}_{(1)}(\omega)=\tilde{f}(\omega)$. From the property of completeness $\sum_{l=1}^{\infty}\tilde{f}_{(l)}(\omega)\tilde{f}_{(l)}^*(\omega')=\delta(\omega-\omega')$, we can derive
\begin{eqnarray}
\sum_{l=1}^{\infty}\hat{a}_j^{\dag}\left[f_{(l)}\right]\hat{a}_k\left[f_{(l)}\right]
&=& \sum_{l=1}^{\infty} \int \tilde{f}_{(l)}(\omega)\hat{\tilde{a}}_j^{\dag}(\omega)\ d\omega \cdot\int \tilde{f}_{(l)}^*(\omega')\hat{\tilde{a}}_k(\omega')\ d\omega' \nonumber \\
&=& \iint \hat{\tilde{a}}_j^{\dag}(\omega)\hat{\tilde{a}}_k(\omega')\sum_{l=1}^{\infty}\tilde{f}_{(l)}(\omega)\tilde{f}_{(l)}^*(\omega')\ d\omega d\omega' \nonumber \\
&=& \int \hat{\tilde{a}}_j^{\dag}(\omega)\hat{\tilde{a}}_k(\omega)\ d\omega.
\end{eqnarray}
Then, the rotation term of $\hat{B}_{j,k}(\kappa,\nu,\mu)$ can be decomposed as follows,
\begin{eqnarray}\label{eq:decomp_bs}
\exp{\left[i\frac{\nu}{2}\hat{L} \right]} &=& \exp{\left[i\frac{\nu}{2}\int \left[\hat{a}_j^{\dag}(\omega)\hat{a}_j(\omega)-\hat{a}_k^{\dag}(\omega)\hat{a}_k(\omega)\right] d\omega \right]} \nonumber \\
&=& \exp{\left[i\frac{\nu}{2}\sum_{l=1}^{\infty} \left[\hat{a}_j^{\dag}\left[f_{(l)}\right]\hat{a}_j\left[f_{(l)}\right]-\hat{a}_k^{\dag}\left[f_{(l)}\right]\hat{a}_k\left[f_{(l)}\right]\right] \right]} \nonumber \\
&=& \otimes_{l=1}^{\infty} \exp{\left[i\frac{\nu}{2} \left[\hat{a}_j^{\dag}\left[f_{(l)}\right]\hat{a}_j\left[f_{(l)}\right]-\hat{a}_k^{\dag}\left[f_{(l)}\right]\hat{a}_k\left[f_{(l)}\right]\right] \right]}.
\end{eqnarray}
The term $\exp{\left[i\frac{\kappa}{2}\hat{M} \right]}$ is also decomposed in the same way. Therefore, $\hat{B}_{j,k}(\kappa,\nu,\mu)$ can be decomposed to a direct product of beam-splitter operations on modes $\left\{ f_{(l)} \right\}_{l=1}^{\infty}$. As for $\hat{D}_j(\alpha,\omega_d)$, we use
\begin{eqnarray}
\sum_{l=1}^{\infty}\tilde{f}_{(l)}^*(\omega_d)\hat{a}_j^{\dag}\left[f_{(l)}\right] 
&=& \int \hat{\tilde{a}}^{\dag}(\omega) \sum_{l=1}^{\infty}\tilde{f}_{(l)}^*(\omega_d)\tilde{f}_{(l)}(\omega)\ d\omega \nonumber \\
&=& \hat{\tilde{a}}^{\dag}(\omega_d).
\end{eqnarray}
From this relation, $\hat{D}_j(\alpha,\omega_d)$ is decomposed as follows,
\begin{eqnarray}\label{eq:decomp_disp}
\hat{D}_j(\alpha,\omega_d) &=& \exp{\left[ \alpha\hat{\tilde{a}}_j^{\dag}(\omega_d)-\alpha^*\hat{\tilde{a}}_j(\omega_d) \right]} \nonumber \\
&=& \otimes_{l=1}^{\infty} \exp{\left[ \left(\tilde{f}_{(l)}^*(\omega_d)\alpha \right)\hat{a}_j^{\dag}\left[f_{(l)}\right]-\left(\tilde{f}_{(l)}^*(\omega_d)\alpha \right)^*\hat{a}_j\left[f_{(l)}\right] \right]}.
\end{eqnarray}
This equation means $\hat{D}_j(\alpha,\omega_d)$ displaces each mode $f_{(l)}$ by the amount of $\tilde{f}_{(l)}^*(\omega_d)\alpha$.

Reference \cite{Yoshikawa2017} shows that for all $f$ we can find an orthogonal mode function $f_{\perp}(t)$ and decompose the photon-pair creation operator $\hat{P}_j^{\dag}[r]$ accordingly,
\begin{eqnarray}
\hat{P}_j^{\dag}[r] =&& ||f^**r_j||\left( \sqrt{M[f,r_j]}\ \hat{a}_j^{\dag}[f]^2+2\sqrt{1-M[f,r_j]}\ \hat{a}_j^{\dag}[f]\hat{a}_j^{\dag}[f_{\perp}] \right)+({\rm irrelevant\ terms}), \nonumber \\
\label{eq:S_decomp}
\end{eqnarray}
where $M[f,r_J] = |\braket{f,N(f^**r_j)}|^2$. The irrelevant terms commute with $\hat{a}_j[f]$. Equation (\ref{eq:S_decomp}) shows that the operator $\hat{S}_j[r_j]$ can be decomposed to a direct product about mode $f$ and irrelevant modes when $M[f,r_j]= 1$. This condition is satisfied when $f(t)$ is a real function and $r_j(t)$ can be regarded as a delta function, that is, the squeezing operation has much broader bandwidth compared to the bandwidth of the filter in the measurement channel. From the assumptions introduced in the Result section, Eq. (\ref{eq:continuous_state}) is given by
\begin{eqnarray}
\ket{\Psi}_1
&\propto& {}_2\bra{\cancel{0}}\hat{a}_2[f]^n\ \hat{U}_{1,2}[f]\otimes\hat{U}_{1,2}\left[\{f_{\perp}\}\right]\ket{\cancel{0}}_1\ket{\cancel{0}}_2 \nonumber \\
&=& {}_2^f\bra{n}\hat{U}_{1,2}[f]\ket{0}_1^f\ket{0}_2^f \otimes \ {}_2^{\{f_{\perp}\}}\bra{0}\hat{U}_{1,2}\left[\{f_{\perp}\}\right]\ket{0}_1^{\{f_{\perp}\}}\ket{0}_2^{\{f_{\perp}\}},
\end{eqnarray}
where $\hat{U}_{1,2}[f]$ and $\hat{U}_{1,2}\left[\{f_{\perp}\}\right]$ are Gaussian operators on  mode $f$ and all modes orthogonal to $f$, respectively. $\hat{U}_{1,2}[f]$ consists of squeezing, beam splitter, displacement operators, and thus the state $\hat{U}_{1,2}[f]\ket{0}_1^f\ket{0}_2^f$ can be an arbitrary Gaussian state by choosing proper parameters $r_1,\ r_2,\ \kappa,\ \nu,\ \mu,\ \omega_d, \alpha_1,$ and $\alpha_2$. Therefore, Eq. (\ref{eq:continuous_state}) is given by
\begin{eqnarray}
\ket{\Psi}_1
&\propto& {}_2^f\braket{n|G}_{1,2}^f \otimes \ket{\chi}_1^{\{f_{\perp}\}},
\end{eqnarray}
which is equivalent to Eq. (\ref{eq:decomp}) and Eq. (\ref{eq:nonGauss}).

{\bf Realization of complex mode function.} 
Let us consider a special case $\hat{U}_{1,2}\ket{\cancel{0}}_1\ket{\cancel{0}}_2=\hat{D}_1(\alpha_1,\omega_d)\hat{D}_2(\alpha_2,\omega_d)\hat{T}_{1,2}[r]\ket{\cancel{0}}_1\ket{\cancel{0}}_2$, where the two-mode squeezing operator $\hat{T}_{1,2}[r]$ is given by
\begin{eqnarray}
&&\ \ \ \hat{T}_{1,2}[r] = \exp{\left( \hat{Q}^{\dag}[r]-\hat{Q}[r] \right)}, \\
&&\hat{Q}^{\dag}[r] = \int \tilde{r}(\omega)\hat{\tilde{a}}_1^{\dag}(\omega)\hat{\tilde{a}}_2^{\dag}(-\omega)\ d\omega.
\end{eqnarray}
The two-mode squeezed state $\hat{T}_{1,2}[r]\ket{\cancel{0}}_1\ket{\cancel{0}}_2$ is also known as an Einstein-Podolsky-Rosen state. We suppose $g(t)\propto f(-t) \in\mathbb{C}$. When $\tilde{r}(\omega)$ is sufficiently broadband than $\tilde{g}(\omega)$, we get the following equation as we prove shortly after
\begin{eqnarray}
\hat{U}_{1,2}\ket{\cancel{0}}_1\ket{\cancel{0}}_2=\hat{U}'_{1,2}[f,f^*]\ket{0}_1^{f}\ket{0}_2^{f^*}\otimes \hat{U}'_{1,2}\left[\{f_{\perp}\},\{f^*_{\perp}\}\right]\ket{0}_1^{\{f_{\perp}\}} \ket{0}_2^{\{f_{\perp}^*\}}.
\label{eq:decomp_U_T}
\end{eqnarray}
Here, $\hat{U}'_{1,2}[f,f^*]$ is a Gaussian operator only acting on mode $f$ of channel 1 and mode $f^*$ of channel 2, and $\hat{U}'_{1,2}\left[\{f_{\perp}\},\{f^*_{\perp}\}\right]$ acts on all other orthogonal modes. Then, the heralded state in mode $f$ on channel 1 is given by
\begin{eqnarray}
\ket{\psi}_1^f =\ {}_2^{f^*}\bra{n}\hat{U}'_{1,2}[f,f^*]\ket{0}_1^{f}\ket{0}_2^{f^*}.
\end{eqnarray}
Thus, we can generate a non-Gaussian state $\ket{\phi}$ in a complex mode $f$.

We derive Eq. (\ref{eq:decomp_U_T}) in the following. The operator $\hat{Q}^{\dag}[r]$ satisfies
\begin{eqnarray}
\left[ \hat{a}_1[f],\hat{Q}^{\dag}[r] \right]
&=&\left[ \int \tilde{f}^*(\omega)\hat{\tilde{a}}_1(\omega)\ d\omega, \int \tilde{r}(\omega')\hat{\tilde{a}}_1^{\dag}(\omega')\hat{\tilde{a}}_2^{\dag}(-\omega')\ d\omega' \right] \nonumber \\
&=& \int \tilde{f}^*(-\omega)\tilde{r}(\omega)\hat{\tilde{a}}_2^{\dag}(\omega)\ d\omega \nonumber \\
&=& \Braket{\tilde{f}^*(-\omega),\tilde{f}^*(-\omega)\tilde{r}(\omega)} \hat{a}_2^{\dag}[f^*]\nonumber \\
&&+ \sqrt{||\tilde{f}^*(-\omega)\tilde{r}(\omega)||^2-|\Braket{\tilde{f}^*(-\omega),\tilde{f}^*(-\omega)\tilde{r}(\omega)}|^2}\ \hat{a}_2^{\dag}[f^*_{\perp}] \nonumber \\
&=& ||f^**r||\left( \sqrt{M'[f,r]}\ \hat{a}_2^{\dag}[f^*]+\sqrt{1-M'[f,r]}\ \hat{a}_2^{\dag}[f^*_{\perp}] \right),
\end{eqnarray}
where $M'[f,r]=|\braket{f^*,N(f^**r)}|^2$. Thus, the decomposition of $\hat{Q}^{\dag}[r]$ corresponding to Eq. (\ref{eq:S_decomp}) is given by
\begin{eqnarray}
\hat{Q}^{\dag}[r] =&& ||f^**r||\left( \sqrt{M'[f,r]}\ \hat{a}_1^{\dag}[f]\hat{a}_2^{\dag}[f^*]+\sqrt{1-M'[f,r]}\ \hat{a}_1^{\dag}[f]\hat{a}_2^{\dag}[f^*_{\perp}] \right)+({\rm irrelevant\ terms}). \nonumber \\
\end{eqnarray}
The irrelevant terms always commute with $\hat{a}_1[f]$. We can show that the irrelevant terms also commute with $\hat{a}_2[f^*]$ when $M'[f,r]=1$ considering decomposition of $\hat{Q}^{\dag}[r]$ based on a commutation relation $\left[ \hat{a}_2[f^*],\hat{Q}^{\dag}[r] \right]$. When $M'[f,r]=1$, $\hat{T}_{1,2}[r]$ is given by a direct product of an operator acting on mode $f$ of channel 1 and mode $f^*$ of channel 2 as well as an operator acting on other modes. Because the decomposition of displacement operation in Eq. (\ref{eq:decomp_disp}) is possible even for complex mode functions, we can decompose $\hat{D}_1(\alpha_1,\omega_d)\hat{D}_2(\alpha_2,\omega_d)\hat{T}_{1,2}[r]$ and get Eq. (\ref{eq:decomp_U_T}).

{\bf Detail of the experiment.} 

\begin{figure}[htbp]
	\begin{center}
		\includegraphics[bb= 0 0 600 180,clip,width=0.6\textwidth]{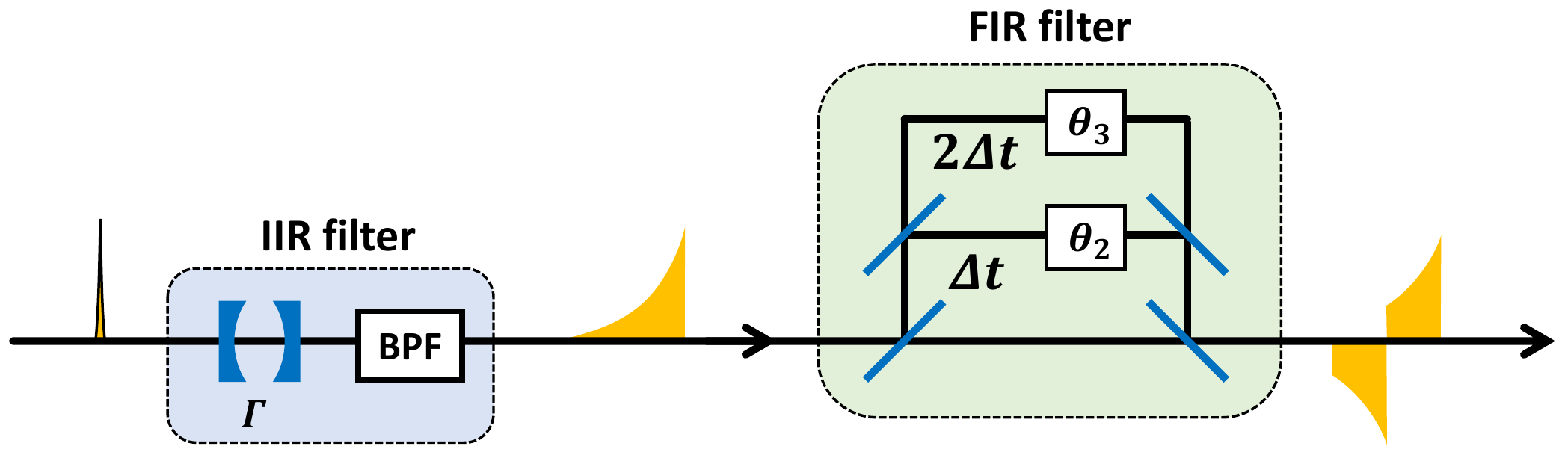}
	\end{center}
	\caption{The diagram of the filter in photon measurement channel. IIR, Infinite Impulse Response; FIR, Finite Impulse Response; BPF, Band Pass Filter.}
\label{Fig:filter}
\end{figure}
Figure \ref{Fig:filter} is the diagram of the filter we used in the experiment. This filter consists of an infinite impulse response (IIR) filter and a finite impulse response (FIR) filter. The IIR filter consists of an optical cavity resonant at $\omega=0$ and a band-pass filter to omit the resonant modes of the cavity at sidebands. Supposing the HWHM of the cavity is $\Gamma$ and the bandwidth of the band-pass filter is sufficiently broad, this step has a low-pass type impulse response given by
\begin{eqnarray}
g_{{\rm IIR}}(t) = 
 \left\{
\begin{array}{ll}
0 & (t < 0)\\
\Gamma \exp{[-\Gamma t]} & (t \ge 0).
\end{array}
\right.
\end{eqnarray}
The FIR filter is an interferometer with three arms. Each arm has a time delay $0, \Delta t,$ and $2\Delta t$. The impulse response of this step is given by
\begin{eqnarray}
g_{{\rm FIR}}(t) = \kappa_1\delta(t) + \kappa_2e^{i\theta_2}\delta(t-\Delta t) + \kappa_3e^{i\theta_3}\delta(t-2\Delta t).
\label{eq:g_if}
\end{eqnarray}
We can realize arbitrary $g_{{\rm FIR}}(t)$ in the form of Eq. (\ref{eq:g_if}) up to a multiplication of a constant. Supposing $\kappa_1:\kappa_2:\kappa_3=1:e^{-\Gamma \Delta t}:0$ and $\theta_1=\pi$, we get the desired impulse response given by
\begin{eqnarray}
g_{{\rm TB}}(t) = 
 \left\{
\begin{array}{ll}
0 & (t < 0,\Delta t \le t)\\
\kappa_1\Gamma \exp{[-\Gamma t]} & (0\le t < \Delta t).
\end{array}
\right.
\end{eqnarray}
We can also realize $g_{{\rm BTB}}(t)$ given by
\begin{eqnarray}
g_{{\rm BTB}}(t) = \left(g_{{\rm IIR}}*g_{{\rm FIR}}\right)(t) = 
 \left\{
\begin{array}{ll}
0 & (t < 0,2\Delta t \le t)\\
\kappa_1\Gamma \exp{[-\Gamma t]} & (0\le t < \Delta t) \\
-\kappa_1\Gamma \exp{[-\Gamma (t-\Delta t)]} & (\Delta t\le t < 2\Delta t),
\end{array}
\right.
\end{eqnarray}
where we put $\kappa_1:\kappa_2:\kappa_3=1:2e^{-\Gamma \Delta t}:e^{-2\Gamma \Delta t}$ and $\theta_2=\pi, \theta_3=0$.

\begin{figure}[htbp]
	\begin{center}
		\includegraphics[bb= 0 0 790 320,clip,width=1\textwidth]{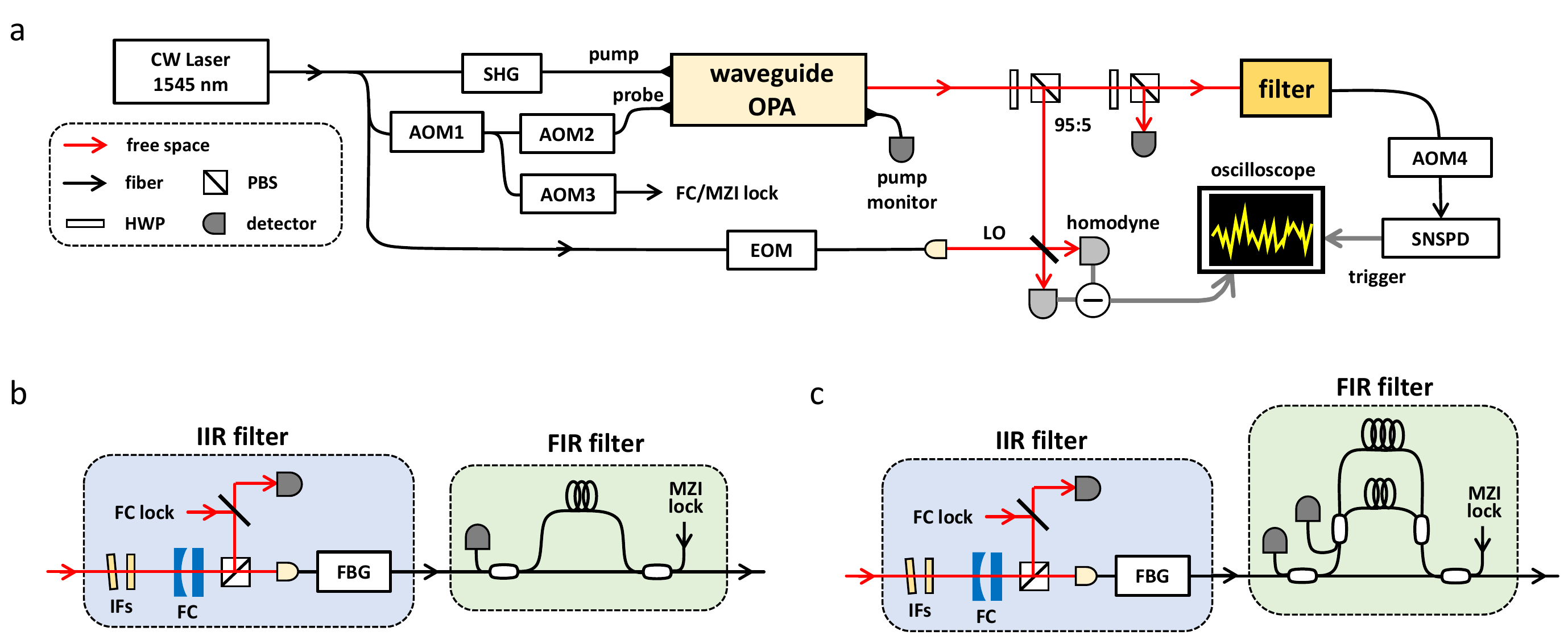}
	\end{center}
	\caption{The detailed diagram of experimental setup. (a) Whole setup. CW, Continuous Wave; AOM, Acousto-Optic Modulator;
SHG, Second Harmonic Generation; EOM, Electro-Optic Modulator; OPA, Optical
Parametric Amplifier; PBS, Polarization Beam Splitter; HWP, Half Wave Plate; LO,
Local Oscillator; SNSPD, Superconducting Nanostrip Photon Detector. (b) The filter for state generation in a time-bin mode. IF, Interference Filter; FC, Filter Cavity; FBG, Fiber Bragg Grating. (c) The filter for state generation in a balanced time-bin mode.}
\label{Fig:setup_TME}
\end{figure}

Figure \ref{Fig:setup_TME} shows the whole setup of the experiment. The main light source is a CW laser at 1545 nm. The CW pump beam is generated at a second-harmonic generation module. The IIR part of the filter consists of a Fabry-P\'{e}rot cavity (HWHM=8.2 MHz, FSR=8.5 GHz), an interferometric filter (HWHM=130 GHz), and a fiber Bragg grating (HWHM=3.6 GHz). The FIR part is a fiber-based interferometer. From $\Delta t=20$ ns and $\Gamma = 2\pi \times 8.2$ MHz, we calculate the ratio of $\kappa_i\ (i=1,2,3)$. We equally distribute the input light to the short, medium, and long arms in the case of the balanced time-bin mode, and to the short and medium arms in the case of the time-bin mode. Then, we adjust $\kappa_i$ to the desired ratio by the variable directional coupler at the output side. The SNSPD is installed into an adiabatic demagnetization refrigerator with an operational temperature around 500 mK. A detection efficiency and a dark count are 63$\%$ and 100 count per second (cps), respectively. We use probe light and lock light as the phase reference of the heralded state and the filter elements, respectively. The phases are locked via piezo elements attached to mirrors and fibers. Probe and lock lights are turned off while the measurement of heralded states because they easily saturate the SNSPD. We use acousto-optic modulators (AOMs) to switch between the control phase and the measurement phase with a period of 1.6 kHz. In the control phase, AOM1 to 3 in Fig. \ref{Fig:setup_TME} is open but the AOM4 is closed. The opposite is true in the measurement phase.

In the time-bin mode experiment, the pump light power is 15 mW, the LO light power 5 mW, and the SNSPD count 21.0 kcps, of which the fake count is 400 cps. In the balanced time-bin mode experiment, the pump light power is 12 mW, the LO light power 10mW, and the SNSPD count 4.2 kcps, of which the fake count is 200 cps. Since the transmittance of the filters is not optimized in this experiment, the state generation rate is not very high especially in the case of the balanced time-bin mode. In principle, it is possible to make a filter with 100$\%$ transmittance for any impulse response \cite{JingujiJan.,JingujiAug.}, so this point can be much improved in the future. The large fake count in the time-bin mode is due to the stray light caused by the LO light. It does not matter in the case of the balanced time-bin mode because the interferometer removes the stray light. The uncertainties of the analysis results are calculated by using bootstrapping method \cite{efron1994introduction}.

\section*{Data availability}
Data will be available from the authors upon request and approval.

\section*{Acknowledgments}
The authors acknowledge supports from UTokyo Foundation and donations from
Nichia Corporation of Japan. W.A. and M.E. acknowledge supports from Research Foundation for OptoScience and Technology. A.K. acknowledges financial supports from The Forefront Physics and Mathematics Program to Drive Transformation (FoPM). This work was partly supported by Japan Society for the Promotion of Science KAKENHI (18H05207, 18H01149, 20J10844, 20K15187) and Japan Science and Technology Agency (JPMJMS2064, JPMJMS2066). The authors would like to thank Mr. Takahiro Mitani for careful proofreading of the manuscript.

\section*{Author contributions}
K.T. conceived and planned the project. K.T., A.K., and B.K.J. designed and constructed the experimental setup and acquired and analyzed the data. Takahiro Kashiwazaki, Takushi Kazama, K.E., K.W., and T.U. supplied the PPLN waveguide OPA. S.M., H.T., M.Y., and F.C. supplied the SNSPD. K.T., W.A., M.E., and J.Y. formulated the theory of the Q-AWG. A.F. supervised the project. K.T. wrote the manuscript with assistance from all other co-authors.

\section*{Competing interests}
The authors declare no competing financial interests.


\bibliography{WPE}

\end{document}